# An explainable XGBoost–based approach towards assessing the risk of cardiovascular disease in patients with Type 2 Diabetes Mellitus

Maria Athanasiou
*School of Electrical and Computer Engineering*
*National Technical University of Athens*
Athens, Greece
mathanasiou@biosim.ntua.gr

Konstantina Sfrintzeri
*School of Electrical and Computer Engineering*
*National Technical University of Athens*
Athens, Greece
konstantinasfr@gmail.com

Konstantia Zarkogianni
*School of Electrical and Computer Engineering*
*National Technical University of Athens*
Athens, Greece
kzarkog@biosim.ntua.gr

Anastasia C. Thanopoulou
*Medical School*
*University of Athens*
Athens, Greece
a_thanopoulou@hotmail.com

Konstantina S. Nikita
*School of Electrical and Computer Engineering*
*National Technical University of Athens*
Athens, Greece
knikita@ece.ntua.gr

*Abstract*—Cardiovascular Disease (CVD) is an important cause of disability and death among individuals with Diabetes Mellitus (DM). International clinical guidelines for the management of Type 2 DM (T2DM) are founded on primary and secondary prevention and favor the evaluation of CVD-related risk factors towards appropriate treatment initiation. CVD risk prediction models can provide valuable tools for optimizing the frequency of medical visits and performing timely preventive and therapeutic interventions against CVD events. The integration of explainability modalities in these models can enhance human understanding on the reasoning process, maximize transparency and embellish trust towards the models' adoption in clinical practice. The aim of the present study is to develop and evaluate an explainable personalized risk prediction model for the fatal or non-fatal CVD incidence in T2DM individuals. An explainable approach based on the eXtreme Gradient Boosting (XGBoost) and the Tree SHAP (SHapley Additive exPlanations) method is deployed for the calculation of the 5-year CVD risk and the generation of individual explanations on the model's decisions. Data from the 5- year follow up of 560 patients with T2DM are used for development and evaluation purposes. The obtained results (AUC=71.13%) indicate the potential of the proposed approach to handle the unbalanced nature of the used dataset, while providing clinically meaningful insights about the model's decision process.

I. INTRODUCTION

Cardiovascular Disease (CVD) is a class of disorders of the heart and blood vessels, that severely affects people with Diabetes Mellitus (DM), being a major cause of disability and death and a barrier to sustainable development. Individuals with Type 2 DM (T2DM), the most common form of DM, are at a substantially increased risk of CVD compared to individuals without DM, experiencing CVD events at an earlier age and presenting higher CVD-related mortality rates [1].

A focus on CVD prevention in high-risk populations can reduce mortality and decrease the imposed economic burden from Coronary Heart Disease (CHD) and stroke. International clinical guidelines for DM management endorse the adoption of cost-effective preventive strategies, which rely on the regular assessment of CVD risk factors for the early detection of high-risk individuals, and trigger targeted interventions in terms of healthy lifestyle behavioural changes and pharmacological treatment prescription [2]. Within this context, computational CVD risk prediction models have great potential to facilitate primary care physicians in accurate CVD risk stratification, enable timely and efficient therapeutic strategies, and eliminate redundant screening procedures for patients at a low CVD risk, thus optimizing patient care pathways and improving quality of life and productivity.

Among a variety of data-driven approaches, survival analysis and regression models have gained the most widespread acceptance in the field of CVD risk prediction models. Such methodologies have been deployed towards the development of the most thoroughly studied CVD risk calculators, including Framingham [3], SCORE [4], and DECODE [5]. Nevertheless, these models underestimate the CVD risk in T2DM individuals due to the underrepresentation of DM patients in the corresponding cohorts [6]. The UKPDS risk engine, dedicated to T2DM patients, has exhibited varying

Part of this research was supported within the framework of ENDORSE project, which is funded by the NSRF. Grant agreement: T1EΔK-03695.

performance, mainly related to the selection criteria of the data underpinning the model, that have resulted in poor reflection of a broader range of T2DM patients [7].

Along these lines, improvements in the performance of computational models predicting health outcomes have been pursued through the application of more sophisticated approaches, investigating techniques in the areas of statistics and machine learning (ML), such as Logistic Regression, Artificial Neural Networks (ANN), Support Vector Machines (SVM), Decision Trees (DT), and Random Forests (RF) [8] [9][10]. More recently, the eXtreme Gradient Boosting (XGBoost) algorithm has attracted significant attention due to its computational speed, generalization capabilities and high predictive performance, and has been employed in various clinical contexts towards CVD risk prediction, chronic kidney disease diagnosis, T2DM onset prediction, and mortality prediction in patients with acute coronary syndrome [11] [12][13]. In the case of DM, the use of ANN, SVM, Bayesian Networks, Self Organizing Maps (SOM), and RF, combined with a limited number of traditional risk factors, has also been investigated for the diagnosis of DM and its long-term complications, including CVD, retinopathy, kidney disease and neuropathy [14] [15] [16] [17].

Despite the promising performance reported in such studies, the models' intricate structure has so far impeded the generation of clinically useful explanations on the reasoning process behind the delivered decisions, thus hampering the adoption of more accurate yet complex risk prediction models in clinical practice. Today, the advent of explainable artificial intelligence opens up new opportunities, offering model-specific and model-agnostic explainability techniques that outweigh any concerns regarding the "black-box" nature of risk calculators, promote transparency and enhance users' trust [18]. In terms of model-agnostic techniques, Local Interpretable Model-agnostic Explanations (LIME) and SHapley Additive exPlanations (SHAP) have been widely used in different domains in order to provide insights about features' contribution in the models' decision process. However, limited explainable risk prediction models have been proposed for assisting disease prognosis and diagnosis [19] [20].

In the present study, an explainable approach, based on the combined use of the XGBoost algorithm and a modification of the SHAP method, i.e., Tree SHAP, is deployed towards the development of an explainable personalized risk prediction model for the fatal or non-fatal CVD incidence in T2DM. A set of well-established CVD-related risk factors are considered for the calculation of a T2DM individual's 5-year risk score to experience a CVD incidence for the first time. XGBoost's computational speed and efficient data handling capabilities are utilized towards exploring nonlinear complex interactions between risk factors. A sophisticated ensemble learning scheme is applied with the aim of addressing the unbalanced nature of the used dataset. A thorough analysis, based on the consistent and computationally efficient Tree SHAP framework, is performed in order to reveal meaningful insights about the ensemble model's decision process and the contribution of the considered risk factors in the final CVD risk scores. To the best of the authors' knowledge, this is the first work proposing an explainable XGBoost-based ensemble learning approach for the calculation of the CVD risk in T2DM.

## II. DATASET

Data collected from a 5-year follow up of 560 T2DM patients at the Hippokration General Hospital of Athens, including 41 patients (7.32%) with CVD incidents during their follow-up period, were used for development and evaluation purposes [15]. Out of the 41 patients, four patients experienced stroke and the rest experienced CHD. Baseline demographic, lifestyle, laboratory and treatment data, that adequately reflect the health status of T2DM patients, were considered in order to compose the model's input space.

## III. METHODOLOGY

Fig. 1 depicts the conceptual framework of the proposed explainable XGBoost-based ensemble approach. It comprises the decision space, featuring the calculation of the 5-year CVD

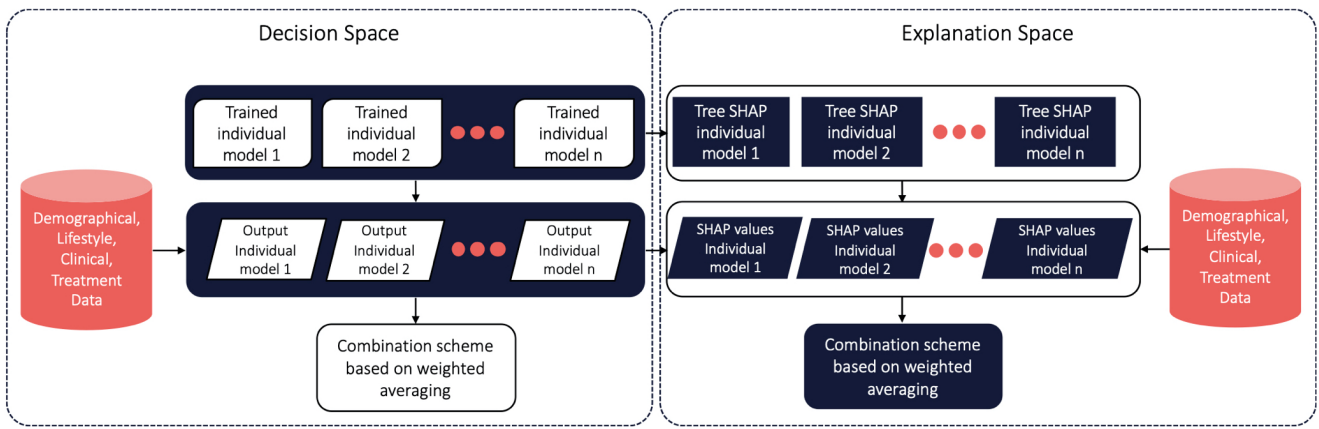

Fig. 1. Conceptual framework. A combination scheme based on weighted averaging was applied in order to merge the individual models' outputs and generate explanations on the ensemble model's decisions.

risk scores, and the explanation space, where explanations on the model's decisions are generated.

*A. Ensemble learning scheme*

*1) Sub-sampling approach:* Considering that training and testing data are usually derived from the same distribution for the development of most prediction models, the unbalanced nature (low number of CVD incidents) of the used dataset would result in over-fitting to the majority (negative for CVD instances) class. In order to avoid over-fitting, an ensemble learning method, based on a sub-sampling approach, was applied towards the generation of appropriate training subsets from the original data. Multiple individual models were trained on these subsets and combined to produce the final CVD risk scores [15]. In particular, the initial training dataset was divided into positive (for CVD) and negative (for CVD) instances. Different sub-samples were subsequently created, in which a 1:2 ratio between the minority (positive instances) and the majority (negative instances) class was preserved, by using all positive instances from the initial training dataset and double sized randomly selected negative instances. In this way, six different training subsets (*n*=6) were derived and used for training six XGBoost-based individual models.

*2) Ensemble model:* The outputs of the trained individual models were combined by means of weighted averaging, as

$$\hat{y}(x) = \frac{1}{n}\sum_{m=1}^{n} w_m \times y_m(x) \quad (1)$$

where $\hat{y}(x)$ is the final risk score in the log-odds scale, $n$ denotes the number of the trained individual models (*n*=6), $w_m$ = [0.15, 0.4, 0.05, 0.1, 0.05, 0.25] and $y_m(x)$ represent the weights and the outputs of the trained models in the log-odds scale, respectively. A brute-force search approach was adopted in order to generate an optimal weights' combination for the six individual models. The final CVD risk scores were obtained by transforming the calculated log-odds scores into the corresponding probabilities, as

$$p(x) = \frac{e^{\hat{y}(x)}}{(1+e^{\hat{y}(x)})}. \quad (2)$$

TABLE I
XGBOOST HYPERPARAMETERS

| Hyperparameter | Value |
|---|---|
| Learning rate | 0.5 |
| Number of estimators | 1000 |
| Minimum split loss | [0.5, 1.0, 0.5, 0.5, 1.0, 2.5] |
| Maximum tree depth | 3 |
| Minimum sum of instance weights in a child | 1 |
| Sub-sample ratio of training instances | 0.8 |
| Sub-sample ratio of columns for each tree | 0.5 |
| Sub-sample ratio of columns for each level | 0.5 |
| L2 regularization term | 2 |
| L1 regularization term | 0 |
| Weight balance between positive-negative class | 4 |

*B. eXtreme Gradient Boosting (XGBoost) algorithm*

eXtreme Gradient Boosting (XGBoost) constitutes an efficient and scalable variant of the Gradient Boosting Machine (GBM) algorithm, leveraging the power of decision tree ensembles towards performance optimization and goodness-of-fit improvement [21]. In XGBoost, multiple decision trees, which are also called weak learners, operate in parallel, and solve both classification and regression problems. GBM utilizes the gradient descent to create a new tree, based on the residual errors of previously generated trees, aiming at the minimization of the objective function

$$Obj = L + \Omega. \quad (3)$$

In (3), $L$ is the training loss function, which measures the model's performance on the training data

$$L = \sum_{i=1}^{N}(y_i - \hat{y}_i)^2, \quad (4)$$

where $y_i$ and $\hat{y}_i$ are the model's target value and output, respectively, and $N$ is the number of instances in the training data. $\Omega$ is a regularization term, which penalizes the complexity of the model in order to avoid over-fitting, and is defined as

$$\Omega = \gamma T + \frac{1}{2}\lambda \sum_{j=1}^{T} w_j^2, \quad (5)$$

where $T$ is the number of leaves, $w_j$ is the weight vector, representing score on the *j*-th leaf, $\lambda$ is L2 regularization and $\gamma$ is a penalization parameter on the number of leaves. The objective function at the *t*-th iteration can be simplified by using the Taylor expansion and fixed term simplifications, and eventually it can be defined as:

$$Obj^{(t)} = \sum_{i=1}^{N}[g_i f_t(x_i) + \frac{1}{2} h_i f_t^2(x_i)] + \Omega(f_t), \quad (6)$$

where

$$g_i = \partial_{\hat{y}^{(t-1)}} loss(y, \hat{y}^{(t-1)}), \quad (7)$$

$$h_i = \partial^2_{\hat{y}^{(t-1)}} loss(y, \hat{y}^{(t-1)}), \quad (8)$$

and $f_t$ represents the tree structure that is added to minimize the objective function.

The identification of the optimal tree structure at each step is performed through a greedy method that enables the selection of the best split candidate for the addition of new leaves, based on the split-induced gain. The gain can be defined as:

$$gain = \frac{1}{2}\left[\frac{(\sum_{i\in I_L} g_i)^2}{\sum_{i\in I_L} h_i + \lambda} + \frac{(\sum_{i\in I_R} g_i)^2}{\sum_{i\in I_R} h_i + \lambda} - \frac{(\sum_{i\in I} g_i)^2}{\sum_{i\in I} h_i + \lambda}\right] - \gamma, \quad (9)$$

where I is the set of instances that are assigned to the leaf before the split, and $I_L, I_R$ are the instance sets of left and right leaf nodes after the split, respectively. This iterative process continues until the tree acquires its defined maximum depth.

*1) Hyperparameters tuning:* In the present study, a brute-force grid-search approach with parallelized performance evaluation was applied towards the optimal tuning of

XGBoost's hyperparameters. Table I summarizes the combination of hyperparameters' values that achieved error minimization in the validation set. This combination was used for tuning the six XGBoost-based individual models. The value of the minimum split loss was differentiated among the individual models towards the generation of tree structures with different levels of complexity. The number of estimators, representing the maximum number of trees that were created during the training phase, was set to 1000. An early stopping criterion was also applied in order to avoid over-fitting, by setting the number of early stopping rounds, i.e., the number of trees, after which tree generation stops unless validation scores continue to improve, to 50.

*2) Feature selection*: Despite the relatively low number of risk factors composing the input space, the importance of the initially considered risk factors was investigated, and an incremental feature selection technique was followed, in order to ensure the selection of the most discriminant and informative features, contributing in beneficial data splits across the models' structure. To this end, feature weight, which is defined as the number of times a feature is used for data splitting, was used as a measure of importance. Average importance values were firstly estimated for each of the 16 features across the six individual models, based on the calculated outputs in the validation set. A hierarchical classification of the features was subsequently performed according to their average importance values, and individual models were trained by using the top-k-ranked features iteratively for k ∈ (1, 2, …, 16). This process was repeated for the different hyperparameters' combinations, generated by grid-search. The top-13-ranked features, depicted in Fig. 2, which achieved the highest AUC score in the validation set, were finally selected to compose the individual models' input space.

*C. Tree SHAP*

SHapley Additive exPlanations (SHAP) constitutes a unified framework, founded on an additive feature attribution method, for generating unique explanation models able to provide explanations on individual models' decisions in the form of particular feature contributions [22]. It is inspired by the concept of shapley values, which in Game Theory describe the players' contribution to the result of a coalitional game, and has gained broad acceptance due to its ability to satisfy local accuracy, missingness and consistency. According to SHAP, a unique explanation model has the form:

$$\varphi_i = \sum_{S \subseteq M \setminus \{i\}} \frac{|S|!\,(M - |S| - 1)!}{M!} [f_{S \cup \{i\}}(x_{S \cup \{i\}}) - f_s(x_S)], \quad (10)$$

where $\varphi_i$ is the shapley value of the *i*-th feature, $M$ is the set of all features, $S$ is a subset of $|S|$ features, $f_{S \cup \{i\}}$ and $f_S$ represent the trained models with the *i*-th feature present and withheld, respectively, and $x_S$ represents the values of features included in the $S$ subset. SHAP approximates the output value $f_x(S)$ as a conditional expectation function of the trained model,

$$f_x(S) = E[f(x)|x_S], \quad (11)$$

where $E[f(x)]$ is the trained model's base value.

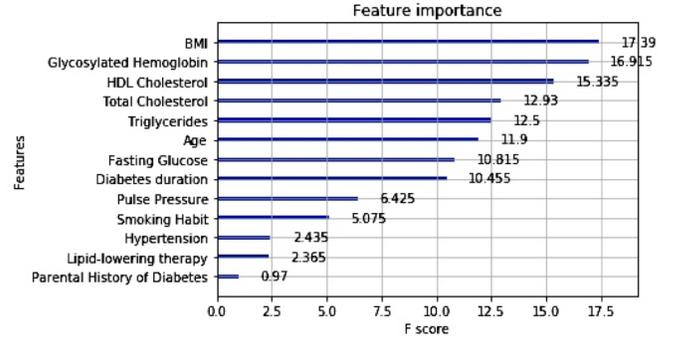

Fig. 2. Average importance values of the top-13-ranked features across the individual models based on feature weight.

In the present study, Tree SHAP, which is a variant of SHAP for tree-based models, characterized by reduced computational complexity, was deployed towards the generation of explanations on the ensemble model's output [23]. To this end, the individual SHAP values of the considered risk factors were calculated for each of the individual models' outputs. Taking into account the inherent linearity of shapley values, a weighted averaging approach, similar to the one applied for obtaining the ensemble model's outputs, was adopted in order to estimate the SHAP values, reflecting the contribution of each risk factor in the final risk scores.

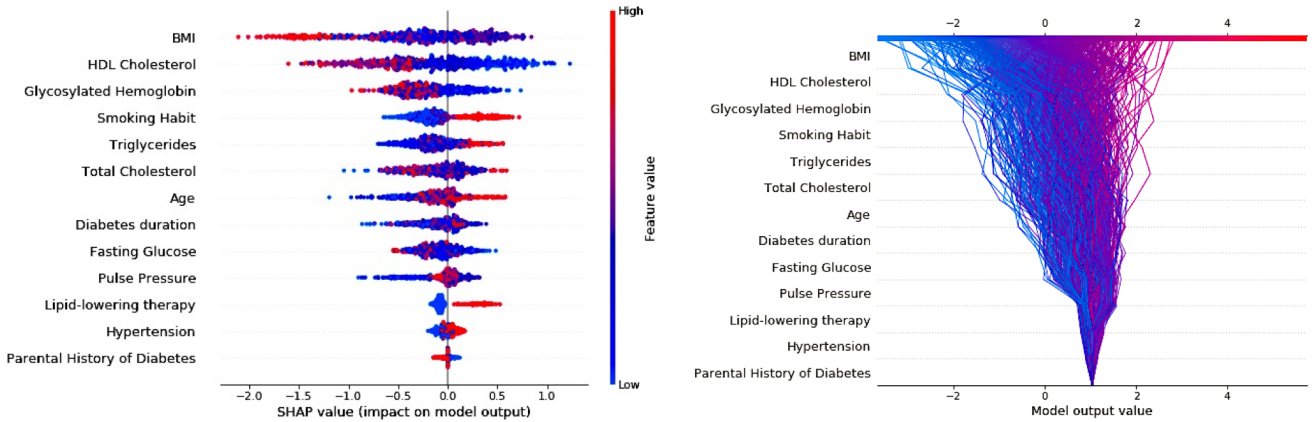

Fig. 3. Left panel: Summary plot depicting the influence of the considered risk factors and their range of effects across the ensemble model's decisions. Risk factors are ranked based on their impact on the model's decisions. Positive and negative SHAP values contribute to the calculation of higher and lower CVD risk scores, respectively. Right panel: Decision plot illustrating the followed prediction paths towards the calculation of CVD risk scores (presented in the log-odds scale) within the model's decision process.

*D. Evaluation framework*

The generalization abilities of the developed risk prediction model were evaluated by applying stratified 10-fold cross-validation, which resulted into 10 data subsets of identical size, approximately preserving the minority representation of the initial dataset. During the training phase of each iteration, training and validation sets were partitioned based on a 90/10 ratio. The ensemble model's predictive performance was assessed in terms of Sensitivity (SENS). The Area Under the Curve (AUC) was employed as a measure of the ensemble model's discrimination ability.

In terms of the obtained SHAP explanations, different visual representations were generated by using the SHAP package (https://github.com/slundberg/shap) in order to capture risk factors' influence and underlying interactions' effects on the ensemble model's outputs. Summary plots were produced to depict the contribution of the considered risk factors and their range of effects across the model's decisions. Decision plots were derived towards identifying potential prediction paths in the model's decision process. Hierarchical clustering analysis was also performed in order to reveal risk factor coalitions towards specific model's decisions [24].

## IV. RESULTS AND DISCUSSION

*A. Evaluation of the model's performance*

Table II summarizes the proposed model's performance, indicating that the ensemble model achieved acceptable discrimination performance (AUC value: 71.13±11.69%). Acceptable sensitivity levels were also obtained, demonstrating the model's ability to handle the unbalanced nature of the used dataset by correctly identifying positive for CVD incidents.

TABLE II
PREDICTIVE PERFORMANCE OF THE PROPOSED RISK PREDICTION MODEL

| Mean ± SD (%) | |
|---|---|
| AUC | SENS |
| 71.13±11.69 | 71.00±23.85 |

*B. Explanations on the model's decisions*

The results of the performed SHAP analysis yielded meaningful insights regarding the impact of the considered risk factors on the ensemble model's decisions. Evidence was also provided on the effectiveness of the proposed approach in delivering explainable CVD risk scores, highlighting its potential to support clinical decision making.

A summary plot, depicting the range of influence of the considered risk factors across the model's outputs, is presented in the left panel of Fig. 3. Colored points were used to illustrate the distribution of the generated decisions' impact among the risk factors. It can be seen that Body Mass Index (BMI), HDL cholesterol and Glycosylated Hemoglobin ($HbA_{1c}$) constituted the most influential risk factors, determining to a great extent the final CVD risk scores across the data instances. Particularly, high levels of BMI and HDL cholesterol and normal $HbA_{1c}$ levels were assigned negative SHAP values, thus contributing in the calculation of lower CVD risk scores. Lower $HbA_{1c}$ levels appeared to contribute in the generation of higher CVD risk scores, despite constituting a protective factor against CVD, which was probably associated with the health profile characteristics of patients in the used dataset. Smoking habit, triglycerides, and older age were also shown to contribute in the increase of the estimated risk scores, while low pulse pressure was attributed negative SHAP values, thus imposing a decrease in the calculated probabilities. Finally, positive SHAP values were generated for lipid-lowering therapy, indicating that the existence of hyperlipidemia disorders led to the calculation of higher risk scores.

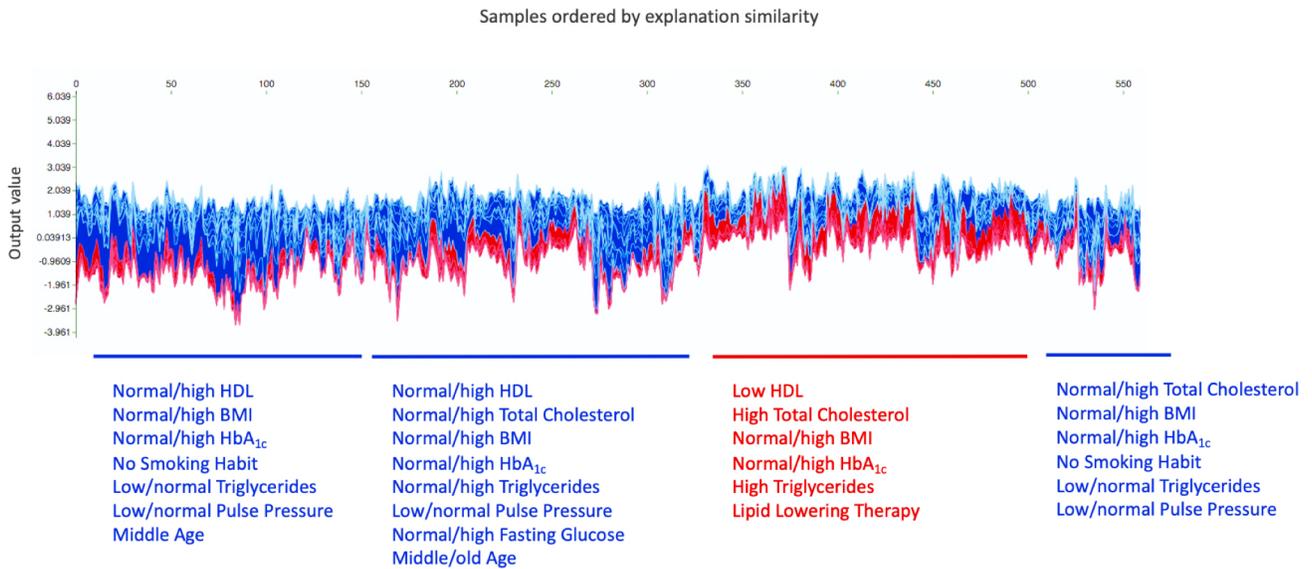

Fig. 4. Clustering of generated decisions based on explanation similarity. Red and blue areas represent risk factors' effects that promote calculation of higher and lower CVD risk scores, respectively. Notable features, that are present in the annotated (through red and blue horizontal lines) clusters, are shown.

A decision plot, summarizing the prediction paths that were followed within the ensemble model's decision process, is presented in the right panel of Fig. 3, where risk factors' SHAP values accumulate towards the final CVD risk scores, depicted in the top horizontal axis. It appears that a common prediction path was followed for the majority of data instances in the beginning of the decision process, featuring risk factors of lower influence such as parental history of diabetes, hypertension and fasting glucose. However, as more influential risk factors were considered, individual prediction paths started to differentiate, evolving towards the calculation of the final risk scores.

Clustering of data instances, influenced by the same coalitions of risk factors, revealed explanation similarities among the generated decisions. The results of hierarchical clustering analysis are depicted in Fig. 4. Notable sets of risk factors were identified, having contributed in the calculation of lower (blue areas) or higher (red areas) CVD risk scores. In particular, normal or high levels of HDL cholesterol, BMI and $HbA_{1c}$, low or normal levels of triglycerides and pulse pressure, and absence of smoking habit, contributed with negative SHAP values, leading to the generation of lower CVD risk scores. On the other hand, positive SHAP values were attributed in the case of co-existence of low HDL cholesterol levels, high levels of total cholesterol and triglycerides, normal or high levels of BMI and $HbA_1$, and lipid lowering therapy, thereby promoting the generation of higher CVD risk scores.

## V. CONCLUSION

Within the framework of the present study, an innovative approach, based on the XGBoost algorithm and the SHAP method, towards the development of an explainable CVD risk prediction model, was proposed. An ensemble learning strategy was adopted and a combination scheme based on weighted averaging was applied to merge the outputs of the trained individual models and the corresponding shapley values of the considered risk factors. The combined use of XGBoost and ensemble learning achieved acceptable performance despite the low number of CVD incidents, demonstrating the method's potential to handle the unbalanced nature of the used dataset.

The obtained explanations revealed clinically meaningful insights about the reasoning process behind the ensemble model's decisions, by capturing the most influential risk factors and highlighting the effect of underlying interactions. Future work concerns the extension of the proposed explainable approach in order to enable the generation of explanations within more advanced ensemble learning strategies, with the ultimate goal to facilitate health professionals in DM patients' risk stratification.

## REFERENCES

[1] International Diabetes Federation, "Diabetes and Cardiovascular Disease," Brussels, Belgium: International Diabetes Federation, 2016.
[2] F. Cosentino *et al.*, "2019 ESC Guidelines on diabetes, pre-diabetes, and cardiovascular diseases developed in collaboration with the EASD," *Eur. Heart J.*, vol. 41, no. 2, pp. 255–323, 2020.
[3] K. M. Anderson, P. M. Odell, P. W. F. Wilson, and W. B. Kannel, "Cardiovascular disease risk profiles," *Am. Heart J.*, vol. 121, no. 1, pp. 293–298, 1991.
[4] R. M. Conroy *et al.*, "Estimation of ten-year risk of fatal cardiovascular disease in Europe: the SCORE project," *Eur. Heart J.*, vol. 24, no. 11, pp. 987–1003, 2003.
[5] B. Balkau *et al.*, "Prediction of the risk of cardiovascular mortality using a score that includes glucose as a risk factor. The DECODE Study," *Diabetologia*, vol. 47, no. 12, pp. 2118-2128, 2004.
[6] R. L. Coleman, R. J. Stevens, R. Retnakaran, and R. R. Holman, "Framingham, SCORE, and DECODE risk equations do not provide reliable cardiovascular risk estimates in type 2 diabetes," *Diabetes Care*, vol. 30, no. 5, pp. 1292–1293, 2007.
[7] A.-P. Kengne *et al.*, "The Framingham and UK Prospective Diabetes Study (UKPDS) risk equations do not reliably estimate the probability of cardiovascular events in a large ethnically diverse sample of patients with diabetes: the Action in Diabetes and Vascular Disease: Preterax," *Diabetologia*, vol. 53, no. 5, pp. 821–831, 2010.
[8] K. V Dalakleidi, K. Zarkogianni, V. G. Karamanos, A. C. Thanopoulou, and K. S. Nikita, "A hybrid genetic algorithm for the selection of the critical features for risk prediction of cardiovascular complications in Type 2 Diabetes patients," in *13th IEEE International Conference on BioInformatics and BioEngineering*, pp. 1–4, 2013.
[9] J. A. Quesada *et al.*, "Machine learning to predict cardiovascular risk," *Int. J. Clin. Pract.*, vol. 73, no. 10, 2019.
[10] K. Dalakleidi, K. Zarkogianni, A. Thanopoulou, and K. Nikita, "Comparative assessment of statistical and machine learning techniques towards estimating the risk of developing type 2 diabetes and cardiovascular complications," *Expert Syst.*, vol. 34, no. 6, 2017.
[11] A. A. Ogunleye and W. Qing-Guo, "XGBoost model for chronic kidney disease diagnosis," *IEEE/ACM Trans. Comput. Biol. Bioinforma.*, 2019.
[12] M. Li, X. Fu, and D. Li, "Diabetes Prediction Based on XGBoost Algorithm," *MS&E*, vol. 768, no. 7, 2020.
[13] K. Antila, N. Oksala, and J. A. Hernesniemi, "High Quality Phenotypic Data and Machine Learning beat a Generic Risk Score in the Prediction of Mortality in Acute Coronary Syndrome," 2019.
[14] M. Skevofilakas, K. Zarkogianni, B. G. Karamanos, and K. S. Nikita, "A hybrid Decision Support System for the risk assessment of retinopathy development as a long term complication of Type 1 Diabetes Mellitus," in *2010 Annual International Conference of the IEEE Engineering in Medicine and Biology*, pp. 6713–6716, 2010.
[15] K. Zarkogianni, M. Athanasiou, A. C. Thanopoulou, and K. S. Nikita, "Comparison of machine learning approaches toward assessing the risk of developing cardiovascular disease as a long-term diabetes complication," *IEEE J. Biomed. Heal. informatics*, vol. 22, no. 5, pp. 1637–1647, 2017.
[16] N. Singh, P. Singh, and D. Bhagat, "A rule extraction approach from support vector machines for diagnosing hypertension among diabetics," *Expert Syst. Appl.*, vol. 130, pp. 188–205, 2019.
[17] P. Dworzynski *et al.*, "Nationwide prediction of type 2 diabetes comorbidities," *Sci. Rep.*, vol. 10, no. 1, p. 1776, 2020.
[18] P. Hall, *An introduction to machine learning interpretability*. O'Reilly Media, Incorporated, 2019.
[19] M. Vega García and J. L. Aznarte, "Shapley additive explanations for NO2 forecasting," *Ecol. Inform.*, vol. 56, 2020.
[20] E. Zihni *et al.*, "Opening the black box of artificial intelligence for clinical decision support: A study predicting stroke outcome," *PLoS One*, vol. 15, no. 4, 2020.
[21] T. Chen and C. Guestrin, "XGBoost," in *Proceedings of the 22nd ACM SIGKDD International Conference on Knowledge Discovery and Data Mining*, pp. 785–794, 2016.
[22] S. M. Lundberg and S.-I. Lee, "A unified approach to interpreting model predictions," in *Advances in neural information processing systems*, pp. 4765–4774, 2017.
[23] S. M. Lundberg *et al.*, "From local explanations to global understanding with explainable AI for trees," *Nat. Mach. Intell.*, vol. 2, no. 1, pp. 2522–5839, 2020.
[24] S. M. Lundberg *et al.*, "Explainable machine-learning predictions for the prevention of hypoxaemia during surgery," *Nat. Biomed. Eng.*, vol. 2, no. 10, pp. 749–760, 2018.